\documentclass[prd,reprint, onecolumn,notitlepage,
               preprintnumbers,showpacs,
               showkeys,superscriptaddress]{revtex4-1}
\usepackage{verbatim}
\usepackage{amsmath, multirow, amssymb, wasysym, gensymb}
\usepackage{graphicx}
\usepackage{units}
\usepackage[utf8]{inputenc}

\begin{document}
\unitlength = 1mm
\title{Novel signatures of the 
       Higgs sector from $\mathsf{S}_3$ flavor symmetry}
\author{Gautam Bhattacharyya}
\email{gautam.bhattacharyya@saha.ac.in}
\affiliation{Saha Institute of Nuclear Physics, 
             1/AF Bidhan Nagar, Kolkata 700064, India}
\affiliation{Fakult\"at f\"ur Physik,
             Technische Universit\"at Dortmund, 
             44221 Dortmund, Germany}
\author{Philipp Leser}
\email{philipp.leser@tu-dortmund.de}
\affiliation{Fakult\"at f\"ur Physik,
             Technische Universit\"at Dortmund, 
            44221 Dortmund, Germany}
\author{Heinrich P\"as}
\email{heinrich.paes@tu-dortmund.de}
\affiliation{Fakult\"at f\"ur Physik,
             Technische Universit\"at Dortmund, 
            44221 Dortmund, Germany}
\preprint{SINP/TNP/2012/07, DO-TH 12/20}
\pacs{14.80.Ec, 11.30.Hv, 12.60.Fr}
\keywords{Higgs boson, flavor symmetry}
\begin{abstract}
In an earlier work we analyzed the $CP$-even scalar sector of an $\mathsf{S}_3$ 
flavor model, where we identified some novel decay signatures of an exotic 
scalar. In this work we extend our analysis by including the complete set of 
scalars/pseudoscalars, revisiting the potential minimization conditions in a 
more general setup, setting the spectrum in conformity with the current LHC 
limits on the scalar mass, and identifying yet another spectacularly novel 
decay channel which might be revealed from an intense study of rare top decays 
at the LHC into modes containing multileptons of different flavors.
\end{abstract}
\maketitle
\section{Introduction}
Discrete flavor symmetries constitute an effective way of explaining the 
masses and mixing of quarks and 
leptons\,\cite{Altarelli:2010gt,*Ma:2007ia,*Ma:2004zd,*Ishimori:2010au}. These 
symmetries may be broken at a high scale by vacuum expectation values 
of scalars, called flavons or familons (see e.g., 
Ref.\,\cite{Feruglio:2009hu}). Interesting experimental signatures such as 
nonstandard decays involving scalars and gauge bosons and/or large flavor 
changing neutral currents often arise in such scenarios.
The flavor group $\mathsf{S}_3$ specifically was introduced early in 
Ref.\,\cite{Pakvasa:1977in} and has since been used in many different flavor 
scenarios\,\cite{Dong:2011vb,*Grimus:2005mu,*Harrison:2003aw,*Ma:1990qh,
*Ma:1999hh,*Mohapatra:1999zr,*Mohapatra:2006pu,*Morisi:2005fy,*Teshima:2005bk,
*Teshima:2012cg,*Teshima:2012cg,*Cao:2011df}. Our analysis is based on the 
flavor symmetry realization introduced in Ref.\,\cite{Chen:2004rr} to describe 
charged lepton and neutrino masses and mixing. The group structure of 
$\mathsf{S}_3$ favors a maximal atmospheric mixing angle which still makes it 
a good fit even after the recent measurements of a nonzero 
$\theta_{13}$\,\cite{Meloni:2012ci,*Dev:2012ns,*Zhou:2011nu}. The group 
$\mathsf{S}_3$ has three irreducible representations: $\mathbf{1}, 
\mathbf{1'}$, and $\mathbf{2}$. The invariants $\mathbf{1}$ can be constructed 
using the multiplication rules 
$\mathbf{2}\otimes\mathbf{2}=\mathbf{1}\oplus\mathbf{1'}\oplus\mathbf{2}$ and 
$\mathbf{1'}\otimes\mathbf{1'}=\mathbf{1}$. We follow the particle 
assignments\,\cite{Chen:2004rr} as we have in Ref.\,\cite{Bhattacharyya:2010hp}:
\begin{equation}
\begin{aligned}
    (L_\mu,L_\tau)&\in \mathbf{2}\, , & L_e, e^c,\mu^c
    &\in \mathbf{1}\, ,
    & \tau^c&\in \mathbf{1'} \, ,\\
    (Q_2,Q_3)&\in\mathbf{2}\, , & Q_1, u^c,c^c,d^c,s^c
    &\in\mathbf{1}\, ,
    & b^c,t^c&\in\mathbf{1'} \, , \\
    (\phi_1,\phi_2)&\in \mathbf{2}\, , & \phi_3&\in\mathbf{1} \,.
\end{aligned}
\end{equation}
The fields $Q_{1/2/3}$ and $L_{e/\mu/\tau}$ refer to the quark and lepton 
$\mathsf{SU}(2)$ doublets of the three generations. This assignment was 
motivated in Ref.\,\cite{Chen:2004rr} in order to have a reasonably successful 
reproduction of quark and lepton masses and mixing. An intuitive appreciation 
of large mixing in the lepton sector \textit{vis-á-vis} small mixing for 
quarks will be clear when we discuss the Yukawa Lagrangian in Sec.~\ref{sec:three}. 
In this paper we concentrate on the scalar $\mathsf{SU}(2)$ doublets 
$\phi_{\{1,2,3\}}$, all of which take part in electroweak symmetry breaking.
The general structure of the model allows for tree-level flavor changing neutral 
currents due to the 
absence of natural flavor conservation\,\cite{Glashow:1976nt}, although those 
are sufficiently suppressed by the Yukawa couplings in this model, and the 
effects remain small even for scalar masses of the electroweak 
scale\,\cite{Bhattacharyya:2010hp,Kubo:2003iw}. However, for models in which 
the Yukawa couplings are not restricted by any flavor symmetry, scalar masses 
are pushed to the TeV scale\,\cite{Yamanaka:1981pa}. In our previous 
analysis\,\cite{Bhattacharyya:2010hp} we found that the additional $CP$-even 
scalars have noteworthy properties. In this work we observe analogous 
properties to hold for pseudoscalars as well. These properties dictate the 
main collider signatures:
\begin{enumerate}
    \item Two of the three scalars $h_{b,c}$ have standard model (SM)-like couplings except
    that they can dominantly decay into the third scalar $h_a$, whose 
    couplings are not SM-like.
    \item The scalar (pseudoscalar) $h_a$ ($\chi_a$) has no 
    $(h_a/\chi_a)VV$-type interactions, where $V\equiv W^\pm,Z$. 
    \item $h_a/\chi_a$ has \emph{only} flavor off-diagonal Yukawa couplings 
    with one fermion of the third generation. 
\end{enumerate}
More specifically, we have extended our previous analysis by including not 
only the $CP$-even neutral scalars, but all scalar degrees of freedom: three 
$CP$-even neutral scalars, two $CP$-odd neutral scalars and two sets of charged 
scalars. \textit{The present analysis extends on our previous 
discussion\,\cite{Bhattacharyya:2010hp} in the following crucial points}: $i$) 
determination of the mass spectrum of the neutral scalars/pseudoscalars and 
the charged scalars following an improved potential minimization technique, 
$ii$) calculation of their couplings to the gauge bosons and matter fields, 
and $iii$) identification of a novel channel of a scalar (pseudoscalar) decay 
within reach of the LHC.
\section{Mass spectrum of the scalars/pseudoscalars}
The general $\mathsf{S}_3$ invariant scalar potential used is given 
by Refs. \cite{Bhattacharyya:2010hp,Kubo:2004ps}
\begin{multline}
    \label{eqn:scalarpotential}
    V = m^2\bigl(\phi_1^\dagger\phi_1+\phi_2^\dagger\phi_2\bigr)
    +m_3^2\phi_3^\dagger\phi_3+\frac{\lambda_1}{2}\bigl(\phi_1^\dagger\phi_1
    +\phi_2^\dagger\phi_2\bigr)^2
    +\frac{\lambda_2}{2}\bigl(\phi_1^\dagger\phi_1
    -\phi_2^\dagger\phi_2\bigr)^2
    +\lambda_3\phi_1^\dagger\phi_2\phi_2^\dagger\phi_1+
    \frac{\lambda_4}{2}\bigl(\phi_3^\dagger\phi_3)^2\\
    +\lambda_5\bigl(\phi_3^\dagger\phi_3\bigr)
    \bigl(\phi_1^\dagger\phi_1+\phi_2^\dagger\phi_2\bigr)
    +\lambda_6\phi_3^\dagger\bigl(\phi_1\phi_1^\dagger
    +\phi_2\phi_2^\dagger\bigr)\phi_3
    +\biggl[\lambda_7\phi_3^\dagger\phi_1\phi_3^\dagger\phi_2
    +\lambda_8\phi_3^\dagger\bigl(\phi_1\phi_2^\dagger\phi_1
    +\phi_2\phi_1^\dagger\phi_2\bigr)+ \text{H.c.}\biggr] \, .
\end{multline}
Once the scalars receive vacuum expectation values, the replacement $\phi_i \to \left(h_i^+, v_i + 
h_i + \mathrm{i}\chi_i\right)^\intercal$ for $i=1\ldots 3$ is performed, with 
the assignments $v_1 = v_2 = v$ and $v_3$, which allow for maximal atmospheric 
mixing. To generate the correct $W^\pm$ and $Z $ masses, 
$2v^2+v_3^2=v_\text{SM}^2$ has to hold, where $v_\text{SM}=246$\,GeV. After 
diagonalizing the mass matrices the masses of the physical 
scalars/pseudoscalars are obtained. These are denoted by 
$h_{a,b,c},\chi_{a,b}$ and $h_{a,b}^+$. 
The remaining degrees of freedom are neutral ($G^0$) and charged ($G^\pm$) 
Goldstone bosons which are eaten up by $Z$ and $W^\pm$ respectively.

Note that $v_1=v_2$ is an extremal point if the following conditions are 
imposed\,\cite{Bhattacharyya:2010hp}:
\begin{equation}
\begin{aligned}
    -m^2 &= (2 \lambda_1+\lambda_3) v^2+ (\lambda_5+\lambda_6+\lambda_7)v_3^2
    +3 \lambda_8 vv_3, & -m_3^2 &= \lambda_4 v_3^2+2
    (\lambda_5+\lambda_6+\lambda_7) v^2 + 2 \lambda_8 v^3/v_3 \, .
\end{aligned}
\end{equation}
To make sure that the extremal point is actually a minimum of the potential, 
the determinant of the Hessian has to be positive. This statement is 
equivalent to imposing the condition of positive squared masses of the 
particles. 
To keep the potential globally bounded from below the conventional approach is 
to arrange all the coefficients of the highest-power terms in the potential to 
be positive definite. This was followed in Ref.\,\cite{Bhattacharyya:2010hp} 
where only the $CP$-even degrees of freedom were considered. However, this 
strategy eliminates the allowed possibility of a large part of valid parameter 
space where the potential is bounded from below although some coefficients 
still stay negative.

Our present analysis is now more complete in the sense that we deal with the 
complete spectrum including all neutral and charged degrees of freedom 
following the potential minimization. Moreover, some parts of the allowed 
parameter space that were \textit{hitherto} cut off by the traditional method 
are now resurrected by our new approach. As a first step, to have an 
analytical feel we identify some simple-looking relations of the coefficients 
by inspection that allow the potential to stay positive and also provide the 
physical scalar masses. To do this the scalar potential in 
Eq.\,(\ref{eqn:scalarpotential}) is factorized into a simplified polynomial in 
$\phi_1,\phi_2$ and $\phi_3$, treating them naively as real quantities for 
calculational ease. There remain three distinct types of terms of order four: 
$\phi_i^4$, $\phi_i^2\phi_j^2$ and $\phi_i^2\phi_j\phi_k$, where 
$i,j,k=1\ldots 3$. Out of the nine terms, only six have independent 
coefficients, called $c_{\{1\ldots 6\}}$:
\begin{equation}
    \label{eqn:exppotential}
    c_1\phi_1^4 + c_1\phi_2^4 + c_2\phi_3^4 + c_3\phi_1^2\phi_2^2 
    + c_4\phi_1^2\phi_3^2 + c_4\phi_2^2\phi_3^2 + 
    c_5\phi_1^2\phi_2\phi_3 + c_5\phi_1\phi_2^2\phi_3 
    + c_6\phi_1\phi_2\phi_3^2\, .
\end{equation}
The coefficients $c_{\{1\ldots 6\}}$ can be expressed in terms of the 
potential parameters $\lambda_{\{1\ldots 8\}}$:
\begin{equation}
\begin{aligned}
    c_1 &= \lambda_1/2 + \lambda_2/2, &c_2 &= \lambda_4/2, &c_3 &= \lambda_1 
    - \lambda_2 + \lambda_3, &c_4 &= \lambda_5 + \lambda_6, &c_5 
    &= 2\lambda_8, &c_6 &= 2\lambda_7\, .
\end{aligned}
\end{equation}
By inspection, we found the following conditions on the coefficients 
$c_{\{1\ldots 6\}}$ from the analytic expressions:
\begin{equation}
\begin{aligned}
    \label{eqn:condbyinspection}
    c_1&>0, & c_2&>0, & 2c_3 &\geq  -c_1, & 2c_3 &\geq  -c_2, 
    & 2c_4&\geq -c_1, & 2c_4&\geq -c_2\, ,\\
    -1/2c_1&\leq c_5\leq c_1, & -1/2c_1&\leq c_6\leq c_1, 
    & -1/2c_2&\leq c_5\leq c_2, & -1/2c_2&\leq c_6\leq c_2\, .
\end{aligned}
\end{equation}
These conditions ensure an acceptable mass spectrum for the neutral 
scalars/pseudoscalars and charged scalars and keep the potential globally 
stable. However, this method renders a large part of the parameter space still 
inaccessible; moreover, the masses obtained by employing 
Eq.\,(\ref{eqn:condbyinspection}) are generally quite light, none exceeding 
$300$\,GeV when ${\left|\lambda_{\{1\ldots 8\}}\right|\leq \pi}$.

To obtain a more complete picture we have transformed 
Eq.\,(\ref{eqn:exppotential}) into spherical coordinates $(\rho,\theta,\phi)$. 
The potential then splits into a radial and an angular part. The question of 
global stability is then reduced to keeping the over-all sign of the angular 
part of the potential positive definite:
\begin{multline}
    \label{eqn:angularpart}
    \sin^4\theta  \bigl\{(2 c_1 -c_3 ) \cos (4 \phi )+6 c_1 +c_3 \bigr\}
    +8 c_2  \cos^4\theta +\sin^2(2 \theta ) \bigl(2 c_4  \sin^2\phi 
    +c_6  \sin (2 \phi )\bigr)\\
    +8 c_4  \cos ^2\phi  \sin^2\theta  \cos ^2\theta 
    +4 c_5  \sin (2 \phi ) \sin ^3\theta  \cos \theta  \bigl(\sin \phi 
    +\cos \phi\bigr)>0 
\end{multline}
\begin{figure}[t!]
    \centering
    \includegraphics[height=2in]{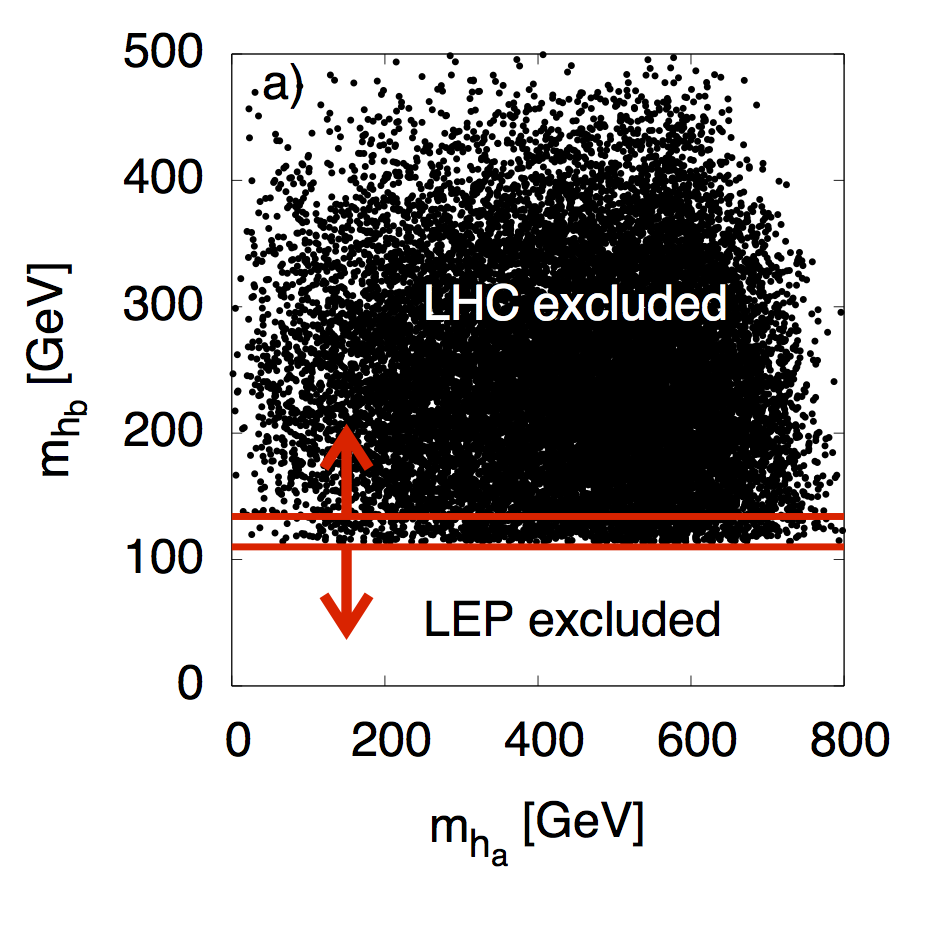}
    \includegraphics[height=2in]{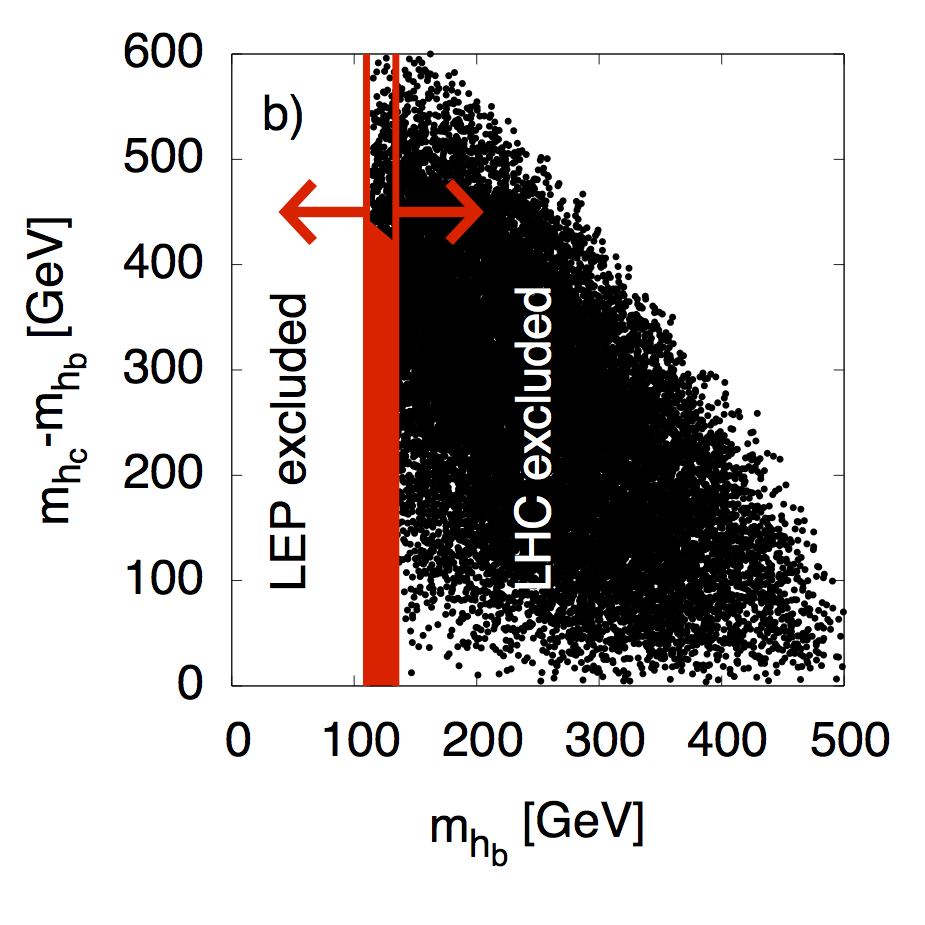}  
    \includegraphics[height=2in]{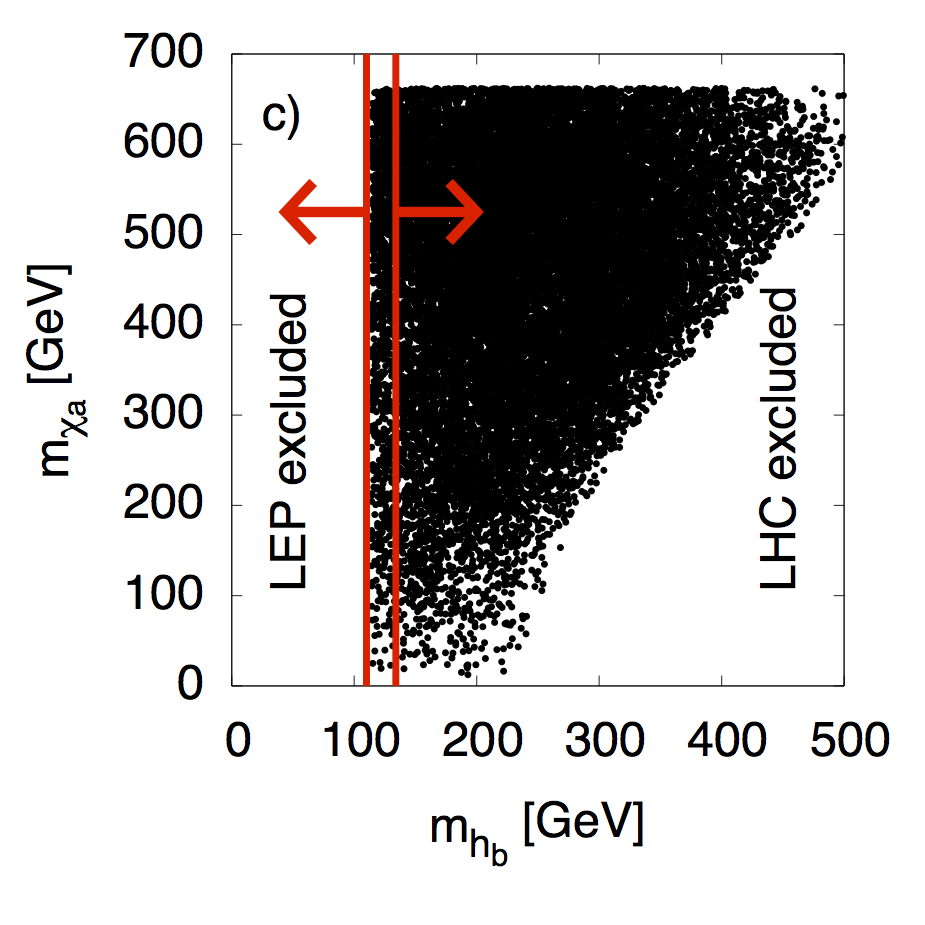}    
    \caption{\label{fig:scatterplots}\textit{Scatter plots of masses of 
    $h_a,h_b,h_c$ and $\chi_a$, where $v_3/v=0.6$. The lines give the current 
    interesting window between $114$\,GeV (LEP2) and $130$\,GeV 
    (LHC)\,\cite{ATLAS-CONF-2012-019,CMS-PAS-HIG-12-008} a) Masses of $CP$-even 
    scalars $h_a$ and $h_b$. b) Mass of $CP$-even scalar $h_b$ plotted against 
    the mass difference of $h_c$ and $h_b$; the highlighted strip is also 
    disfavored by LHC which rules out a second SM-like Higgs within 
    $550$\,GeV. c) Mass of $h_b$ compared to that of the $CP$-odd scalar 
    $\chi_a$.}}
\end{figure}
As this is a transcendental inequality there is no analytically tractable and 
simple set of conditions that can be imposed on the coefficients $c_{\{1\ldots 
6\}}$ to solve Eq.\,(\ref{eqn:angularpart}). We therefore decided to check the 
positivity of this function numerically at each point of the parameter space. 
This allows us to explore the so-far inaccessible territory of the stable 
parameter space that could not be reached by the conditions of 
Eq.\,(\ref{eqn:condbyinspection}). Consequently, masses well beyond $300$\,GeV 
for the scalars/pseudoscalars can be reached even keeping 
$\left|\lambda_{\{1\ldots 8\}}\right|\leq\pi$. To sum up, our 
Eq.\,(\ref{eqn:condbyinspection}) is an improvement over what we have done in 
Ref.\,\cite{Bhattacharyya:2010hp} and subsequently our numerical approach 
improves the size of the accessible parameter space even further. 

Diagonalizing the mass matrix of the pseudoscalars gives the symmetry basis 
pseudoscalars $\chi_{1,2,3}$ in terms of the physical basis pseudoscalars 
$\chi_{a,b}$:
\begin{equation}
\begin{aligned}
    \label{eqn:pseudoscalarmixing}
    \chi_{1(2)} &= \left(v/v_\text{SM}\right) G^0 \mp \left(1/\sqrt{2}\right) 
    \chi_a - v_3/\left(\sqrt{2} v_\text{SM}\right)\chi_b, 
    & \chi_3 &= \left(v/v_\text{SM}\right) G^0 
    + \sqrt{2}\left(v_3/v_\text{SM}\right)\chi_b\, .
\end{aligned}
\end{equation}
It is interesting to note that the mixing coefficients are very simple and 
just depend on the ratio $v_3/v_\text{SM}$. This is in stark contrast to the 
mixing of $h_{1,2,3}$ and $h_{a,b,c}$ given in 
Ref.\,\cite{Bhattacharyya:2010hp}, where the coefficients are complicated 
functions of the $\lambda_{\{1\ldots 8\}}$ parameters of 
Eq.\,(\ref{eqn:scalarpotential}):
\begin{equation}
\begin{aligned}
 \label{eqn:scalarmixing}
 h_{1(2)}&= U_{1(2)b}~h_b + U_{1(2)c}~h_c \mp \left(1/\sqrt{2}\right) ~h_a, & 
 h_3&= U_{3b}~h_b + U_{3c}~h_c \, , 
\end{aligned}
\end{equation}
where $U_{ib}$ and $U_{ic}$ are complicated
functions of the $\lambda_{\{1\ldots 8\}}$, $v$ and $v_3$. The mixing 
relations for the charged scalars $h_{a,b}^+$ are obtained by substituting 
$\chi\to h^+$ and $G^0\to G^+$ in Eq.\,(\ref{eqn:pseudoscalarmixing}). The 
masses for the $CP$-even scalars are\,\cite{Bhattacharyya:2010hp}
\begin{equation}
\begin{aligned}
    m_{h_a}^2 &= 4\lambda_2 v^2 - 2\lambda_3 v^2 - v_3
    \left(2\lambda_7 v_3 + 5\lambda_8 v\right)\, ,\\
    m_{h_{b(c)}}^2 &= \frac{1}{2v_3}\left[4 \lambda_1 v^2 v_3+2
    \lambda_3 v^2 v_3+2 \lambda_4 v_3^3-2 \lambda_8 v^3+3
    \lambda_8 v v_3^2 \mp \Delta m^3\right]\, ,
    \label{eqn:msqscalars}
\end{aligned}
\end{equation}
where 
\begin{multline}
    \label{eqn:delta-m3}
    \Delta m^3 = \biggl[8 v v_3 \biggl\{2 v v_3^3 \biggl(2
    (\lambda_5+\lambda_6+\lambda_7)^2-\lambda_4 (2
    \lambda_1+\lambda_3)\biggr)+2 \lambda_8 v^4 (2 \lambda_1+\lambda_3)-3
    \lambda_4 \lambda_8 v_3^4\\
    +12 \lambda_8 v^2 v_3^2
    (\lambda_5+\lambda_6+\lambda_7)+12 \lambda_8^2 v^3
    v_3\biggl\}
    +\biggl\{2 v^2 v_3 (2 \lambda_1+\lambda_3)+2
    \lambda_4 v_3^3-2 \lambda_8 v^3+3 \lambda_8 v
    v_3^2\biggr\}^2\biggr]^\frac{1}{2}\, .
\end{multline}
We now derive the pseudoscalar squared masses as
\begin{equation}
\begin{aligned}
    m_{\chi_a}^2 &= -9 \lambda_8 v v_3, & m_{\chi_b}^2 &= -v_\text{SM}^2 
    \left(2\lambda_7 + \lambda_8 v/v_3\right)\, .
\end{aligned}
\end{equation}
The corresponding squared masses for the charged scalars are
\begin{equation}
\begin{aligned}
    m_{h_a^+}^2 &= -2\lambda_3 v^2 - v_3^2 \left(\lambda_6 + \lambda_7 \right) 
    + 5 \lambda_8 v v_3, 
    & m_{h_b^+}^2 &= -v_\text{SM}^2\left(\lambda_6+\lambda_7 
    + \lambda_8 v/v_3\right)\, .
\end{aligned}
\end{equation}

The allowed ranges for the masses can be found by a random scattering in the 
parameter space where the couplings in the potential are varied within 
$\lambda_{\{1\ldots 8\}}\in[-\pi,\pi]$ and the ratio $v_3/v$ is fixed to $0.6$ 
(this value was chosen in Ref.\,\cite{Bhattacharyya:2010hp} for compatibility 
with the quark masses). The allowed range for the couplings 
$\lambda_{\{1\ldots 8\}}$ has been increased with respect to what was assumed 
in our previous work\,\cite{Bhattacharyya:2010hp} to admit a broader mass 
spectrum. The values are still very much within pertubative bounds. This leads 
to a $CP$-even mass spectrum similar to Ref.\,\cite{Bhattacharyya:2010hp}, but 
with higher allowed ranges. The scalar $h_a$ can be as massive as roughly 
800\,GeV or arbitrarily light as it evades the LEP2 bound due to its maximally 
nonstandard couplings. The mass of the SM-like scalar $h_b$ is limited within 
$114$--$500$\,GeV, while $h_c$ is still heavier. Both $h_b$ and $h_c$ masses 
should however satisfy the LEP2 lower bound of $114$\,GeV [See 
Fig.~\ref{fig:scatterplots} for details].

In view of the recent LHC 
results\,\cite{ATLAS-CONF-2012-019,CMS-PAS-HIG-12-008} that hint towards a 
SM-like Higgs boson at around $125$\,GeV with a large excluded region above 
and below, the mass spectrum in this model is compatible with the following 
scenario:
\begin{enumerate}
    \item $h_b$ plays the role of the SM-like Higgs boson with a mass of 
    roughly $125$\,GeV. The Yukawa and gauge couplings of $h_b$ and $h_c$ are 
    SM-like with numerically negligible flavor off-diagonal 
    couplings\,\cite{Bhattacharyya:2010hp}.
    \item The scalars/pseudoscalars $h_a$ and $\chi_a$ have nonstandard 
    interactions that hide them from standard searches, as will be discussed 
    in the following sections. In particular, $h_a$ and $\chi_a$ can be very 
    light.
    \item All other scalar/pseudoscalar masses, including the charged 
    scalars, can be above $550$\,GeV. We however note that the existing limits 
    on charged scalar masses are not so stringent and the parameters of our 
    potential can be arranged to admit a much smaller mass for them, though 
    this is not the main focus of our present work.
\end{enumerate}
\section{Couplings of the scalars/pseudoscalars}
\label{sec:three}
\begin{table}[h]
    \begin{center}
        \begin{tabular}{c||c|c|c|c|c|c}
            & $\mathbf{h_a^\pm W^\mp}$ & $\mathbf{h_b^\pm W^\mp}$ 
            & $\mathbf{\chi_aZ}$ & $\mathbf{\chi_bZ}$ & $\mathbf{W^\pm W^\mp}$  
            & $\mathbf{ZZ}$\\
            \hline\hline
            $\mathbf{h_a}$ & $\checkmark$ & -- & $\checkmark$ &--&--&--\\
            $\mathbf{h_b}$ &-- &$\checkmark$ & --&$\checkmark$ & $\checkmark$ 
            & $\checkmark$\\
            $\mathbf{h_c}$ &-- &$\checkmark$ &-- &$\checkmark$ & $\checkmark$ 
            & $\checkmark$
        \end{tabular}\quad\quad
        \begin{tabular}{c||c|c}
            & $\mathbf{h_a^\pm W^\mp}$ & $\mathbf{h_b^\pm W^\mp}$\\
            \hline\hline
            $\mathbf{\chi_a}$ & $\checkmark$ &--\\
            $\mathbf{\chi_b}$ &  -- & $\checkmark$\\
            &
        \end{tabular}
    \end{center}
    \caption{\label{tab:gaugecouplings}\textit{Three-point vertices involving 
    at least one neutral scalar/pseudoscalar and gauge bosons. A checkmark 
    indicates that the vertex exists.}}
\end{table}
It is worth noting that among the couplings listed in 
Table \ref{tab:gaugecouplings} the ones involving $h_a$ do not depend on any 
parameters of the scalar potential, while the couplings of $h_b$ and $h_c$ to 
the gauge bosons are complicated functions of the scalar mixing parameters, 
which we refer to in our tables by putting checkmark signs without displaying 
their explicit forms. The $h_a\chi_aZ$ coupling has a simple form
 $h_a\chi_aZ: -\frac{\mathrm{i}}{2} G q_\mu$,
where $G=\sqrt{g^2+g'^2}$ and $q_\mu$ is the momentum transfer. As stated in 
Ref.\,\cite{Bhattacharyya:2010hp}, $h_a$ stands out because it does not couple 
to pairs of gauge bosons via the three-point vertex. As a result, neither the 
LEP2 lower limit of $114$\,GeV nor the electroweak precision test upper limit 
of around $200$\,GeV applies to it. The same is true for the pseudoscalar 
$\chi_a$.
For certain kinematic regions, the coupling $h_a\chi_a Z$ is important for 
collider searches as we shall see later.
Table \ref{tab:othercouplings} contains the other gauge-scalar-scalar and the 
triple-scalar vertices.
\begin{table}[h]
    \begin{center}
        \begin{tabular}{c||c|c|c|c}
            & $\mathbf{h_a^\mp  \gamma}$ & $\mathbf{h_a^\mp Z }$ &  
            $\mathbf{h_b^\mp  \gamma}$ & $\mathbf{h_b^\mp Z }$\\
            \hline\hline
            $\mathbf{h_a^\pm}$ &
            $\checkmark$ & $\checkmark$ & -- &--\\
            $\mathbf{h_b^\pm}$ &
            --&-- &$\checkmark$ & $\checkmark$\\
            &&&&
        \end{tabular}\quad\quad
        \begin{tabular}{c||c|c|c|c|c|c|c|c|c}
            & $\mathbf{h_{a}h_{a}}$& $\mathbf{h_ah_b}$ 
            & $\mathbf{h_ah_c}$&$\mathbf{h_{a}^\pm 
            h_{a}^\mp}$&$\mathbf{h_{b}^\pm h_{b}^\mp}$
            &$\mathbf{h_{a}^\pm h_{b}^\mp}$
            &$\mathbf{\chi_{a}\chi_{a}}$
            &$\mathbf{\chi_{b}\chi_{b}}$
            &$\mathbf{\chi_{a}\chi_{b}}$\\
            \hline\hline
            $\mathbf{h_a}$ &--& $\checkmark$ & $\checkmark$ & -- 
            &--&$\checkmark$ &-- & --&$\checkmark$\\
            $\mathbf{h_b}$ & $\checkmark$ &-- &-- &$\checkmark$ & $\checkmark$ 
            &--& $\checkmark$ & $\checkmark$&--\\
            $\mathbf{h_c}$ & $\checkmark$ &-- &-- &$\checkmark$ & $\checkmark$ 
            &-- &$\checkmark$ & $\checkmark$&--
        \end{tabular}
    \end{center}
    \caption{\label{tab:othercouplings}\textit{Other three-point vertices. A 
    checkmark indicates that the vertex exists.}}
\end{table}
Note that $h_a$ couples only off-diagonally to the other 
scalars/pseudoscalars. The $h_ah_a^\pm h_b^\mp$ couplings depend only on 
$v_3/v$, while the other triple scalar couplings are complicated functions of 
the scalar mixing parameters.

For illustration, we first write the Yukawa Lagrangian of the $CP$-even scalars 
in the basis $\{h_1,h_2,h_3\}$ with the couplings $f_i$ for leptons and $g_i$ 
for quarks as in Ref.\,\cite{Bhattacharyya:2010hp}:
\begin{multline}
    \mathcal{L}_{\text{Yuk}} = f_4 e e^c h_3 + f_5 e \mu^c h_3 
    + f_1 \mu^c (\mu h_2 + \tau h_1) + f_2 \tau^c (- \mu h_2 + \tau h_1)
    +g_4^u u u^c h_3 + g_5^u u c^c h_3 
    + g_1^u c^c (c h_2 + t h_1)\\ + g_2^u t^c (- c h_2 + t h_1)  
    + g_4^d d d^c h_3 + g_5^d d s^c h_3 
    + g_1^d s^c (s h_2 + b h_1) + g_2^d b^c (- s h_2 + b h_1)
    + \text{H.c.}
\end{multline}
We then rotate the scalars in the Yukawa Lagrangian to their physical basis 
$\{h_a,h_b,h_c\}$ which gives the Yukawa matrices $Y_{\{a,b,c\}}$.
The individual mixing matrices for up- and down-type quarks contain large 
angles as a consequence of $S_3$ symmetry and the particle 
assignments\,\cite{Chen:2004rr}. Specifically, the doublet representation of 
$S_3$ generates maximal mixing when $v_1=v_2$. Now, the 
Cabibbo-Kobayashi-Maskawa matrix involves a relative alignment of those two 
matrices which yields small mixing for quarks. Similarly, the 
Pontecorvo-Maki-Nakagawa-Sakata matrix is given by the relative orientation of 
the mixing matrices for the charged leptons and neutrinos. Since the neutrino 
mass matrix generated in the present context by a type-II seesaw mechanism 
turns out to be diagonal, the large mixing angles in the lepton sector 
survive. There are two generic textures of Yukawa couplings in this 
model\,\cite{Bhattacharyya:2010hp}:
\begin{equation}
\begin{aligned}
    Y_{a} &= \begin{pmatrix}
                0 & 0 & Y_{13} \\
                0 & 0 & Y_{23} \\
                Y_{31} & Y_{32} & 0
             \end{pmatrix},
    & 
    Y_{b,c} &= \begin{pmatrix}
                Y_{11} & Y_{12} & 0\\
                Y_{21} & Y_{22} & 0\\
                0 & 0 & Y_{33}
               \end{pmatrix}\, .
\end{aligned}
\end{equation}
Here $Y_{a}$ symbolically describes the Yukawa couplings for $h_a,\chi_a$ and 
$h_a^+$, while $Y_{b,c}$ describe the couplings for $h_b, h_c, \chi_b$ and 
$h_b^+$. The pattern holds both for leptons and 
quarks\,\cite{Bhattacharyya:2010hp} and reproduces the observed masses and 
mixing\,\cite{Chen:2004rr}. The off-diagonal couplings in $Y_{b,c}$ are 
numerically small and can be controlled by one free parameter which keeps 
processes like $\mu\to e\gamma$ and meson mixing well under control. The 
largest off-diagonal coupling in $Y_{a}$ is $(h_a/\chi_a) c t$ which is about 
$0.8$; it leads to viable production channel of $h_a$ via $t$ decays as 
described in the next section. The next largest couplings are $(h_a/\chi_a) s 
b \approx 0.02$ and $(h_a/\chi_a) \mu \tau \approx 0.008$. The $\chi_a \mu 
\tau$ coupling induces an interesting decay channel potentially observable at 
the LHC. Note that since the $h_a^+ t b$ coupling does not exist the mass of 
$h_a^+$ is not constrained by the LHC searches in the $t\to h^+b$ channel in 
the mass window of $80$--$160$\,GeV\,\cite{:2012cw,Aad:2012tj}.

\section{Observing $h_a$ at the LHC}
\begin{figure}[t!]
    \centering
    a) \includegraphics[height=1in]{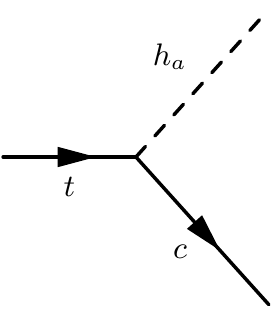}\quad\quad\quad
    b) \includegraphics[height=1in]{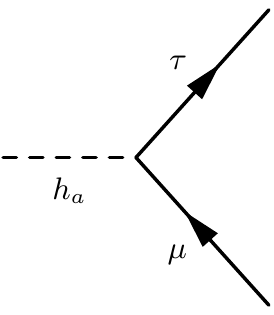}\quad\quad\quad 
    c) \includegraphics[height=1in]{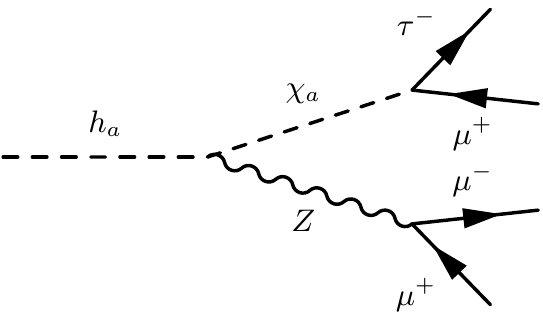}
    \caption{\label{fig:feynmangraphs}\textit{Feynman graphs for dominant 
    sources of $h_a$ production and decays which might be relevant at the LHC.
    }}
\end{figure}

If kinematically allowed the dominant production of $h_a$ occurs through $t\to 
h_a c$ [Fig.~\ref{fig:feynmangraphs}(a)]. The subsequent decay channels depend 
crucially on the mass of the pseudoscalar $\chi_a$: if $m_{h_a}<m_{\chi_a}$, 
$h_a$ decays dominantly into $b$ and $s$ quarks, or $\tau$ and $\mu$ [see 
Fig.~\ref{fig:feynmangraphs}(b)]. The branching ratio (BR) for $t\to h_ac$ is 
about $0.17 (0.06)$ for $m_{h_a}= 130 (150)$\,GeV. Then $h_a\to\mu\tau$ 
proceeds with a BR of $10\%$ and $h_a\to b s$ with $90\%$.
\begin{figure}[t!]
    \centering
    \includegraphics[height=2in]{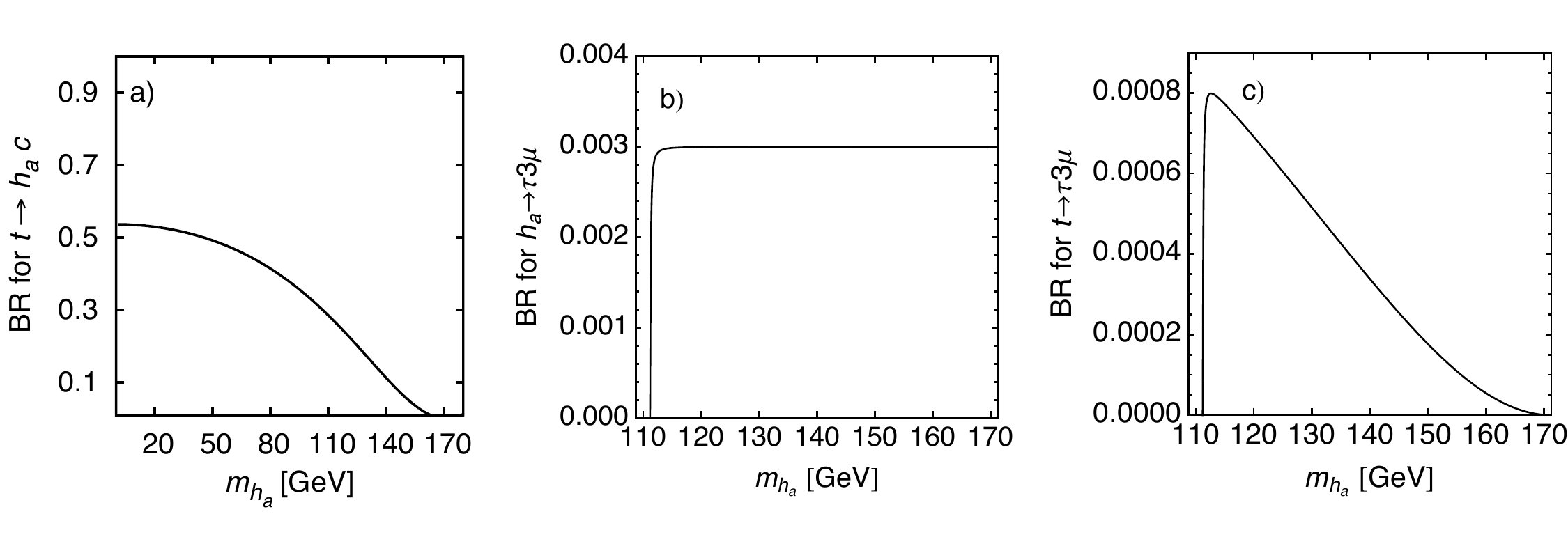}
    \caption{\label{fig:hadecay}\textit{Different branching ratios involving 
    the production and decay of $h_a$. In all cases, $m_{\chi_a}=20$\,GeV is 
    assumed.
    }} 
\end{figure}

A spectacular channel opens up when $h_a\to \chi_a Z$ is kinematically 
accessible [Fig.~\ref{fig:feynmangraphs}(c)]. The BR of $h_a\to Z\chi_a$ is 
almost $100\%$ due to the numerical dominance of the gauge coupling over the 
Yukawa couplings involving light fermions, followed by $\chi_a\to\tau\mu$ with 
a BR of $\sim 10\%$, and a $Z\to \mu\mu$ BR of $\sim 3\%$. If two $h_a$ are 
produced from $t\bar{t}$ pairs, this could lead to a characteristic signal 
with up to six muons with the tau tags. The BRs for $t\to c h_a$ and 
subsequently $h_a\to\chi_a Z \to\tau\mu\mu\mu$ are plotted in 
Figs.~\ref{fig:hadecay}(a)--(c). For these plots $m_{\chi_a}=20$\,GeV has been 
assumed, which is allowed by current data. The BR peaks for $m_{h_a}=110$\,GeV 
once the kinematic threshold is crossed and then falls sharply for larger 
masses due to phase space constraints.

\section{Conclusions}
This is a natural extension of our previous work\,\cite{Bhattacharyya:2010hp} 
where we studied only the scalar sector assuming the pseudoscalars to be too 
heavy to be relevant. In this work we have analyzed the complete 
scalar/pseudoscalar sector of an $\mathsf{S}_3$ flavor model. We deal with 
three $CP$-even, two $CP$-odd and two sets of charged scalar particles. In this 
work we have improved our potential minimization technique which enabled us to 
explore a larger region of the allowed parameter space. It is possible to 
arrange the mass spectrum in full compatibility with the current LHC data, 
with the scalar $h_b$ mimicking the Higgs-like object lurking around 
$125$\,GeV. The specific scalar (pseudoscalar) with prominent non-standard 
gauge and Yukawa interactions, namely $h_a$ ($\chi_a$), evade standard 
searches at LEP/Tevatron/LHC and hence can be rather light. The other 
scalars/pseudoscalars can be arranged to stay beyond the current LHC reach 
(e.g., $550$\,GeV). In particular, we have identified a promising channel for 
$h_a$ search involving up to six muons in the final state with the tau tags. 
We urge our experimental colleagues to dig out this information which is 
probably buried in the existing data.
\begin{acknowledgments}
G.B.~acknowledges DFG support through a Mercator visiting professorship, grant 
number INST 212/289-1, and hospitality at TU Dortmund. P.L.~was supported by the 
Studienstiftung des deutschen Volkes, and both G.B.~and P.L.~acknowledge 
hospitality at CERN PH/TH where this work was initiated. H.P.~was supported by 
DFG grant PA 803/6-1. We thank J.~Kamenik for his valuable comments.
\end{acknowledgments}

\end{document}